# Optical bandgap and bowing parameter for Fe doped LaGaO$_3$

P.R. Sagdeo*, Preetam Singh, Hari Mohan Rai, Rajesh Kumar
Material Research Lab. (MRL), Department of Physics and MSE; Indian Institute of Technology Indore, Simrol, Indore–452020, India.

Archna Sagdeo,
Indus Synchrotrons Utilization Division, Raja Ramanna Center for Advance Technology (RRCAT), Indore–452013, India.

Parasmani Rajput
Atomic & Molecular Physics Division, Bhabha Atomic Research Centre, Trombay, Mumbai-400085, India.

**Abstract:** The polycrystalline samples of LaGa$_{1-x}$Fe$_x$O$_3$ have been prepared by standard solid state reaction route. The phase purity of the prepared samples is confirmed by powder x-ray diffraction experiments followed by Rietveld analysis. It has been observed that the variation of lattice parameters is governed by Vegard's law. The optical band gap of these samples is estimated using diffuse reflectance analysis (DRA) and it is observed that the optical gap systematically decreases with Fe doping from 3.62 eV and attains the saturation value of ~1.9 eV at x= 0.4. The value of the bowing parameter 'b' for the prepared solid solution LaGa$_{1-x}$Fe$_x$O$_3$ is estimated to be 3.8eV. The x-ray absorption near edge spectroscopy (XANES) suggests that the Fe is in mixed valence state in all prepared samples and these mixed states of Fe due to off-stoichiometry acts like electron doping in LaGa$_{1-x}$Fe$_x$O$_{3\pm\delta}$ and thereby results in the reduction in the effective band gap. Our results may be useful to design the LaGaO$_3$ based light emitting diodes and new generation of semiconductor photo-detectors.

*Corresponding author: prs@iiti.ac.in

**Introduction:** The design and the development of direct band gap materials for light emitting diodes (LED), field emission display and florescent displays etc. are of great scientific and technological interest. In this connection the indium gallium arsenide (InGaAs) is one of the most successful candidates[1].For InGaAs and similar semiconductor materials the relation between the band gap and lattice parameters (bowing parameter) has been reported[2]. One of the main problems associated with the InGaAs based structure is degradation due to oxidation (also termed as ageing effect)[3–5]. In this connection the direct band gap oxides have attracted much attention[6–10]. The $LaGaO_3$ is well known wide band gap perovskite oxide material having band gap for bulk sample close to 4.0 eV[11,12]. Importantly $LaGaO_3$ has been widely explored for the field emission applications by doping of Yb, Eu, Ce, Cr Sm, Tb, Tm etc. and known to exhibit excellent photo luminescent properties[6,13–15]. However, the systematic study such as effect of doping concentration, dopants, etc.on the structural properties, optical properties and the relationship between them considering Vegard's law and bowing parameter etc. for doped $LaGaO_3$ is hardly explored. It is important to notice here that the end compound $LaGaO_3$ and $LaFeO_3$ and solid solution i.e. intermediate compound possesses orthorhombic symmetry with space group *Pnma*[16–18]. Hence the Fe doped $LaGaO_3$ may be ideal solid solution to study the optical band gap, bowing parameter and Vegard's law. It should be noted that due to difference in the ionic radius of $Ga^{+3}$ and $Fe^{+3}$ it is expected that with Fe doping in $LaGaO_3$may lead to systematic variation in lattice parameters as observed in similar perovskites oxides and consequently variation in band gap[2,9,10,19]. Hence, in order to understand the effect of Fe doping on the structural and optical properties; in the present study we have doped Fe in $LaGaO_3$ at Ga site and prepared the polycrystalline samples of $LaGa_{1-x}Fe_xO_3$ by solid state reaction route. The diffraction studies followed by Rietveld analysis indicate that the variation of lattice parameters in $LaGa_{1-x}Fe_xO_{3\pm\delta}$series is governed by Vegard's law. The diffuse reflectance analysis (DRA) suggests that the optical band gap in the studied samples varies from 3.62 eV to 1.90 eV with the value of the bowing parameter 'b' is to be 3.8 eV.

**Experimental:** Polycrystalline samples of LaGa$_{1-x}$Fe$_x$O$_3$ (0≤x≤1) were prepared by conventional solid-state reaction route[20] with starting materials; La$_2$O$_3$ (99.99%), Ga$_2$O$_3$ (99.99%), and Fe$_2$O$_3$ (99.98%). These starting materials were mixed in stoichiometric amount and homogeneously mixed with Propanol as a mixing medium. The resulting homogenous mixture was calcined in air ambient at 1000 °C, 1110°C, and 1200 °C each time for 24 hours and final sintering was carried out at 1350°C in air for 24 hours. In order to examine the phase purity of the prepared samples the powder x-ray diffraction (XRD) experiments were carried out on Bruker D8 diffractometer equipped with Cu target having LYNX EYE detector. The optical band gap of prepared samples has been measured using diffuse reflectivity measurements. These measurements have been performed in the 190 nm to 800 nm wavelength range using Cary-60 UV-VIZ-NIR spectrophotometer having Harrick Video-Barrelino diffuse reflectance probe. The beam spot size on the sample was around 1.5 mm in diameter and an integral sphere detector is used for diffuse signal detection. Fe K-edge XANES measurements have been carried out on the scanning extended X-ray absorption fine structure (EXAFS) beamline [21](BL-9) of Indus-2 synchrotron source in fluorescence mode using Vortex detector. The energy range of XANES was calibrated using Fe foil.

**Results and discussion:**

Figure-1shows the powder XRD pattern for prepared samples the representative Rietveld refined diffraction data for LaGa$_{0.8}$Fe$_{0.2}$O$_3$is shown in figure-2. The diffraction data shown in the figure-1 shows the systematic shift towards lower 2θ value with Fe doping, the same is shown in the inset of figure-2. In order to estimate the value of lattice parameters with Fe doping we have refined the obtained diffraction data for all prepared samples considering orthorhombic structure with *Pnma* space group[17]. The value for the goodness of fit was found to be close to 1.40 for all samples. It should be noted that the absence of any unaccounted peak in the refined XRD patterns confirms the phase purity of the prepared samples.Figure-3 shows the variation of lattice

parameters as a function of Fe doping. The increase in the value of lattice constant can be understood in the terms of difference in the ionic radius of $Ga^{+3}$ (0.62) and $Fe^{+3}$ i.e. (0.645)[22]. There exists a linear variation of lattice parameters with Fe doping; suggesting that the solid solution $LaGa_{1-x}Fe_xO_3$ follows the Vegard's law[23] i.e.

$$\mathbf{a}(LaGa_{1-x}Fe_xO_3) = \mathbf{a}(LaGaO_3)*(1-x) + \mathbf{a}(LaFeO_3)*(x) \ldots\ldots\ldots\ldots\ldots\ldots(1)$$

Here $\mathbf{a}(LaGa_{1-x}Fe_xO_3)$ is the lattice parameter for $x$ fraction of Fe doped sample; whereas $\mathbf{a}(LaGaO_3)$ and $\mathbf{a}(LaFeO_3)$ are the lattice parameters for pure $LaGaO_3$ and $LaFeO_3$ respectively. In order to confirm the same we have estimated the lattice parameters of intermediate samples using the lattice parameters of $LaGaO_3$ and $LaFeO_3$ by Vegard's law and compared the same with the experimentally observed lattice parameters; the same is shown in figure-3. The one to one match between the observed and calculated value of lattice parameter suggests that the solid solution $LaGa_{1-x}Fe_xO_3$ follows Vegard's law. Further from the figure 3 it is clear that the value of lattice parameter "$a$" both experimental and as derived from Vegard's law matches well as compared to that of $b$ and $c$; this may be due to the possible room temperature orbital ordering of $Fe^{+3}$ and $Fe^{+4}$ (or higher charge state of Fe) ions in $b$-$c$ plane in the prepared samples[24–26] or due to the possible Jahn-Teller effect associated with Fe ion[27] this need further investigations. It should be noted that the near edge x-ray absorption spectroscopy studies (as discussed in next section) confirms the presence of Fe in mixed valence state in the prepared samples. Further Vegard's law assumes isotropic/spherical expansion of unit cell but due to non-spherical shape of the atomic orbitals in most of the cases this assumption may not hold true.

The optical bandgap of the prepared samples has been determined using diffuse reflectance spectroscopy by converting the diffuse reflectance in to equivalent absorption spectra by using the Kubelka–Munk function of the following form:

$$F(R_\infty) = \frac{K}{S} = \frac{(1-R\infty)^2}{2R\infty} \ldots\ldots\ldots\ldots\ldots(2)$$

Here $R\infty = R_{Sample}/R_{Standard}$. $R_{Sample}$ is the diffuse reflectance of the sample and $R_{Standard}$ is that of the standard ($BaSO_4$ in present case). K and S are the Kubelka–Munk absorption and scattering functions, respectively. If the material scatters in a perfectly diffuse manner, the scattering function S is nearly constant with wavelength[28] and the Kubelka–Munk function can be related/proportional to the absorption coefficient (α) as

$$F(R_\infty) \propto \alpha \propto \frac{(h\nu - Eg)^{1/n}}{h\nu} \ldots\ldots\ldots\ldots\ldots\ldots(3)$$

Here n has the value of 2 for direct bandgap transitions, while n is equal to 1/2 for an indirect bandgap transition. Thus, a plot between $[F(R\infty) \times h\nu]^n$ versus $h\nu$ yields a straight line and the intercept on the energy axis gives the value of the bandgap. The value of n = 2 has been taken to determine the optical gap of $LaGa_{1-x}Fe_xO_3$ solid solutions as the doped $LaGaO_3$ is known to demonstrate excellent photoluminescence properties (which is a possible signature of direct band gap material) [6,13–15]. Fig. 4 shows a plot between $[F(R\infty) \times h\nu]^2$ and $h\nu$ for $LaGa_{1-x}Fe_xO_3$. From the figure it is clear that with increase in the doping of Fe in $LaGaO_3$ the intensity feature at E = 3.32 eV and E = 2.7 eV become more prominent. It should be noted that the x-ray diffraction data confirms the phase purity of the prepared samples, hence; the intensity feature at E = 3.32 eV and E = 2.7 eV appears to be intrinsic property of the Fe doped $LaGaO_3$. The occurrence of new states at lower energy with doping may be understood either in the terms of electron/hole doping or due to defects (vacancies) present in the sample[29–31]. In order to understand the occurrence of new low energy states in the Fe doped samples we have carefully examined the valency of Fe in $LaGa_{1-x}Fe_xO_3$ using XANES[32,33]. Figure-5 show the XANES spectra for the studied samples along with that of FeO ($Fe^{+2}$) and $Fe_2O_3$ ($Fe^{+3}$) standards. As discussed above the energy axis is calibrated using pure Fe foil as standard and the value of the absorption edge is estimated by taking the derivative of the edge region [34]. It should be noted that the absorption edge of pure Fe, FeO and $Fe_2O_3$ are found to be at 7112 eV, 7120 eV and 7125 eV respectively; whereas the absorption edge for the Fe doped $LaGaO_3$ samples ranges between 7127.3 eV to 7128.5 eV. The edge jump value is at higher side for Fe K-edge (as clear from XANES data) in Fe doped LaGaO$_3$ samples suggests that the fraction of Fe is present in oxidation state greater than +3 in the prepared samples[32,33]. The presence of Fe in higher oxidation state may be termed as electron doping due to non-stoichiometric nature of Fe doped LaGaO$_3$ samples. This electron doping may give rise to the defects states at lower energy levels[29–31] and may be responsible for the observed intensity features at E = 3.32 eV and E= 2.7 eV as observes in case.

Figure-6 shows the variation of optical band gap Eg in Fe doped LaGaO$_3$ as function of Fe doping. From the figure-6 it is clear that with increase in Fe doping in LaGa$_{1-x}$Fe$_x$O$_3$ the optical band gap gradually decreases and attains the saturation value of ~1.9 eV at x = 0.4. The decrease in the value of band gap can be understood in the terms of many body effects associated with the orbital overlap (bond length, bond angles etc.), interaction of free charge carries with ionized lattice points (effective mass m*) and defects states [28-30,34,35]. The data shown in the figure 6 is fitted to estimate the value of bowing parameter using equation[37]

$$Eg\ [_{LaGa1-xFexO3}] = (1-x) * [Eg:_{LaGaO3}] + x * [Eg:_{LaFeO3}] - b *(x) * (1-x)\ldots\ldots..(4)$$

Here, Eg: LaGaO$_3$ and Eg:LaFeO$_3$ are the optical gaps of LaGaO$_3$ and LaFeO$_3$ and b is the bowing parameter. The value of band gap obtain for LaFeO$_3$ is in good agreement with the values reported in the literature[37]. We have fitted the experimental optical gap data using equation (4) as shown by the dotted line in figure 6. The best fit was obtained for a bowing parameter value of +3.8 ± 0.05 eV. The positive value of bowing parameter may be attributed to a repulsive interaction between the unoccupied extended conduction band states and the occupied donor-like d-levels of Fe[37,38] or due to the possible orbit coupling at Γ point of zone boundary[39] this needs further investigations. Moreover, the appearance of extra states (as shown in figure 4) points towards the presence of extra electron (or hole) in the prepared samples and these states appears in between VBM (valence band maxima) and CBM (conduction band maxima) and these possible states are shown in a schematically proposed band diagram as illustrated in the inset of figure 6. These states are possibly arising due to the coexistence of $Fe^{3+}$ and $F^{4+}$ in the studied samples[40]. These defects states may also be related to the Urbach tail states this needs further investigations by taking in to account actual values of absorption coefficient $\alpha$[41].

Further it will be interesting to study the photoluminescence properties of these samples.

**Conclusion:** A systematic study has been carried out to investigate the possible correlations between optical gap and lattice parameter possibly for the first time for technologically important $LaGaO_3$ by means of Fe doping. It is observed that the variation of lattice parameters for $LaGa_{1-x}Fe_xO_3$ follows the Vegard's law. The optical gap in $LaGa_{1-x}Fe_xO_3$ systematically decreases with Fe doping and attains the saturation value of ~1.9 eV at x= 0.4. The XANES study shows that the Fe is in mixed valence state. Our results may be useful to design the $LaGaO_3$ based light emitting diodes and new generation of semiconductor photo-detectors.

**Acknowledgement:** Authors sincerely thank Prof. P. Mathur Director IIT Indore for his encouragement. CSIR New Delhi is acknowledged for funding the high temperature furnace under the project 03(1274)/13/EMR-II used for the sample preparations.Authors thank RRCAT Indore for providing beamtime for x-ray spectroscopy experiments. One of the authors Mr. Preetam Singh likes to thank Mr. Ravikiran Late for his help during x-ray spectroscopy experiment. Mr. Preetam Singh is thankful to Mr.Vikas Mishra for his help in DRA measurements and data analysis and Mr. Shailendra K. Saxena, Mr. Kamal Warshi and Mrs. Pryanka for encouragement. SIC IIT Indore is acknowledged for providing basic infrastructure facilities.

# Figure Captions

***Figure 1:*** *Powder x-ray diffraction pattern for the prepared Fe doped LaGaO$_3$ samples.*

***Figure2:*** *Rietveld refined powder x-ray diffraction data for LaGa$_{0.8}$Fe$_{0.2}$O$_3$. The insets shows the magnified views for (121) at 2θ = 32.45 and (442) at 2θ = 95.44. The inset on the right hand side shows the shift in 2θ values for pure and Fe doped LaGaO$_3$ samples.*

***Figure 3:*** *Variation of lattice parameters as a function of Fe doping.*

***Figure 4:*** *The equivalent Tauc plot for LaGa$_{1-x}$Fe$_x$O$_3$ samples.*

***Figure5:*** *Representative XANES spectra for the prepared Fe doped LaGaO3 samples.*

***Figure 6:*** *Variation of band gap in LaGa1-xFexO$_3$ with increasing Fe. The dotted line shows the fitting using equation 4.*

**FIGURE-1**

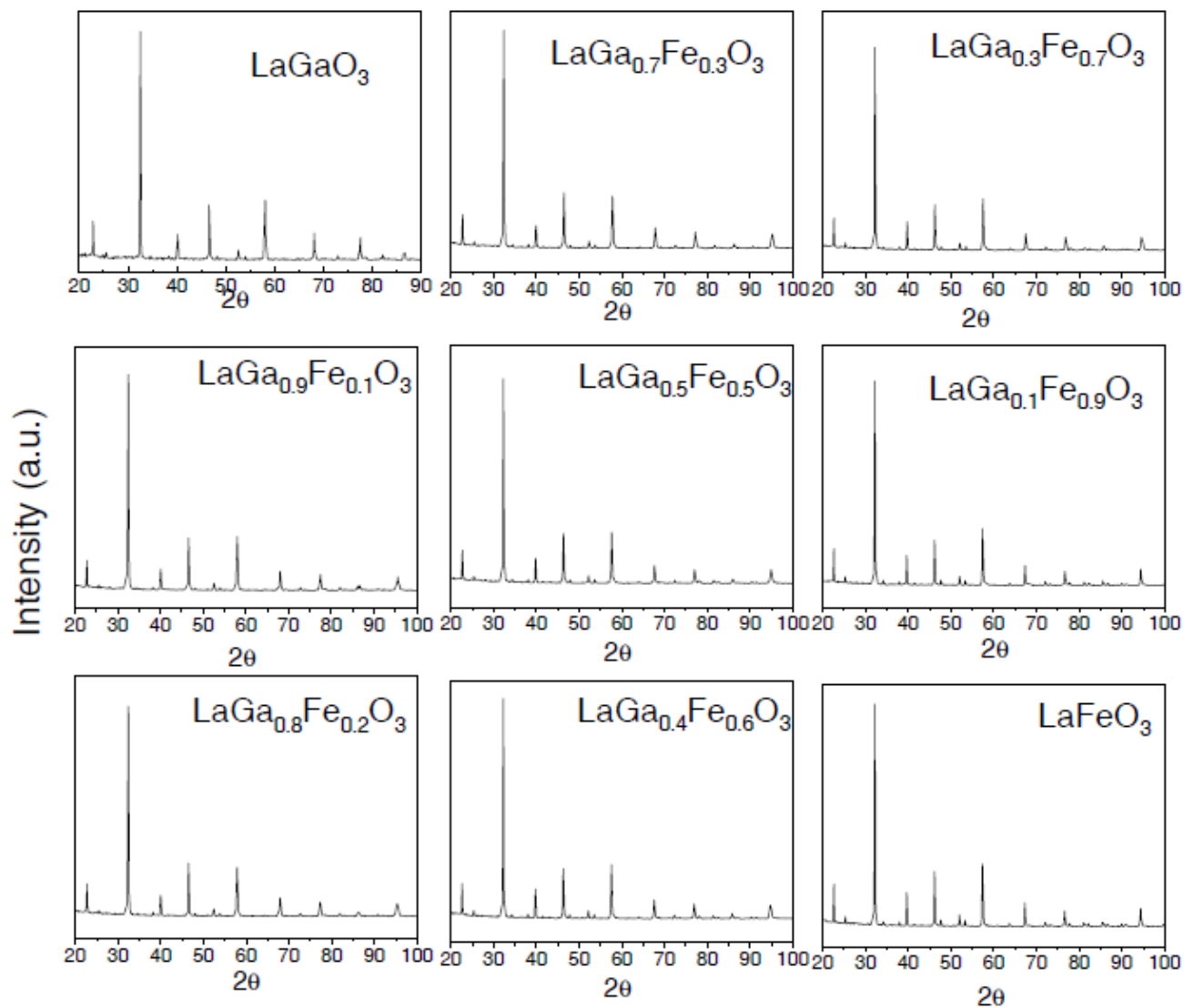

Figure 2

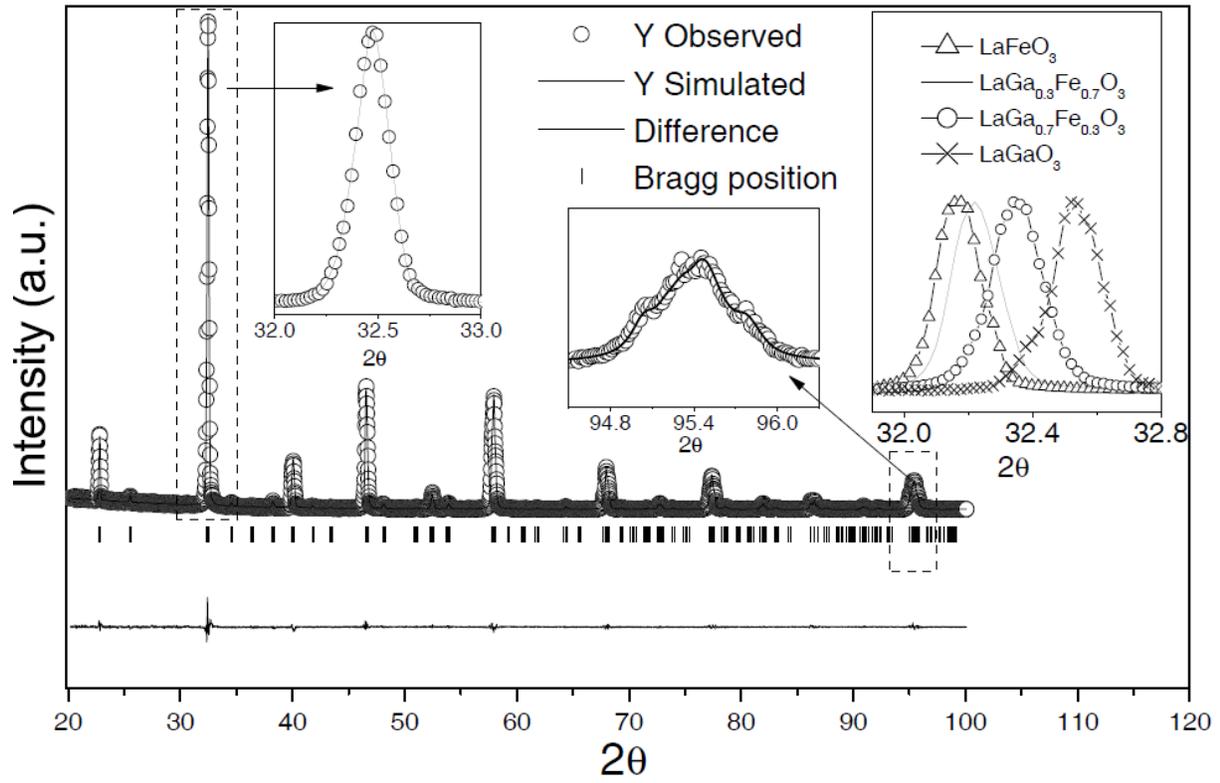

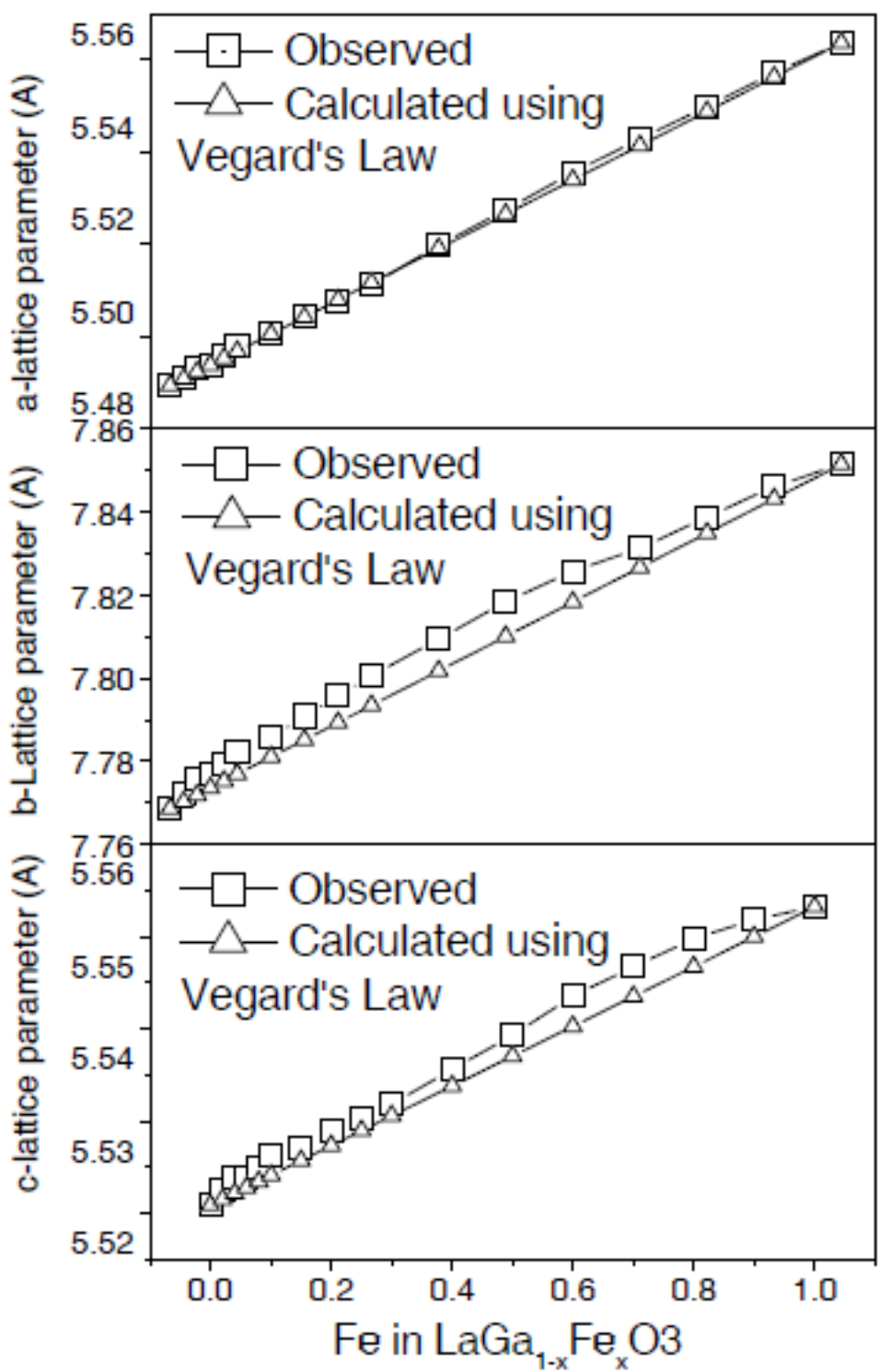

Fe in LaGa$_{1-x}$Fe$_x$O3

Figure 4

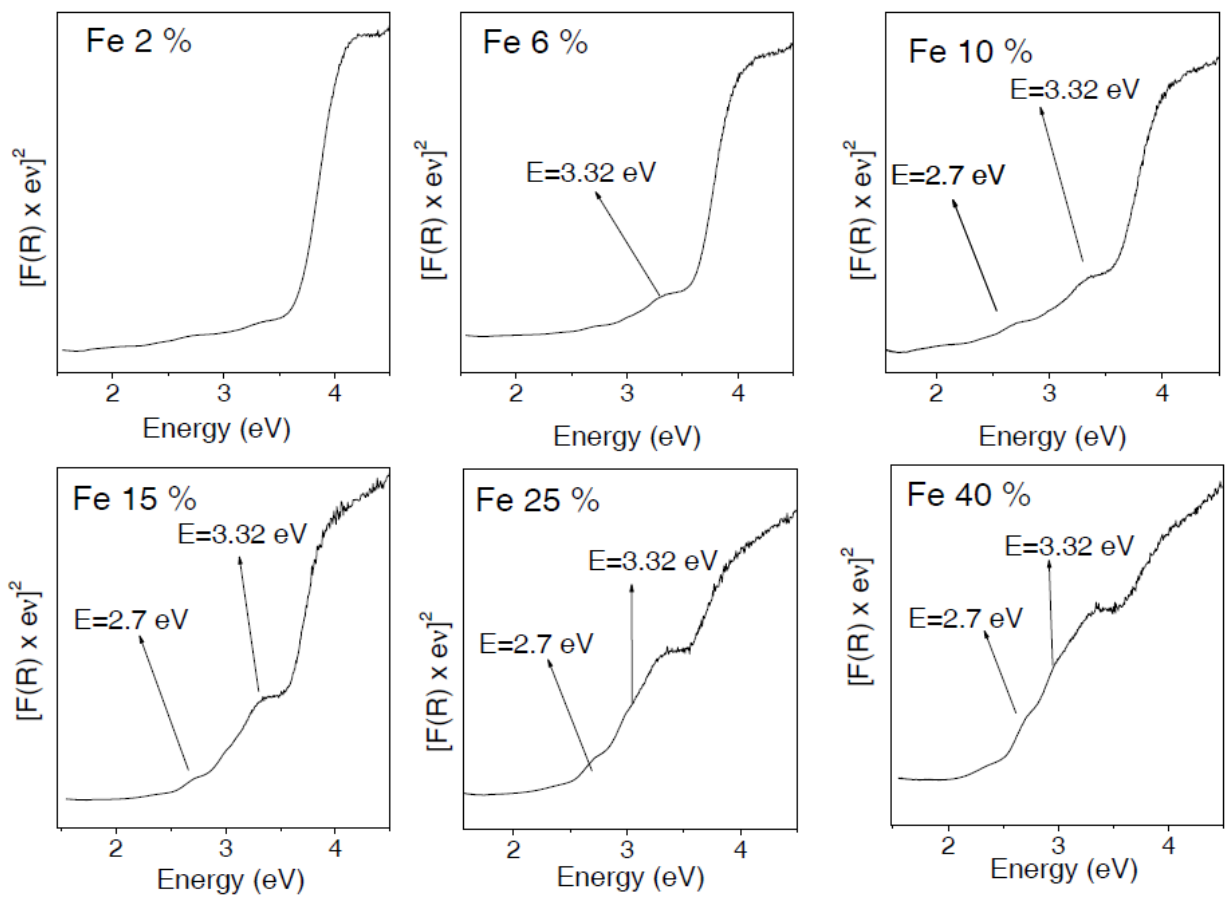

Figure 5

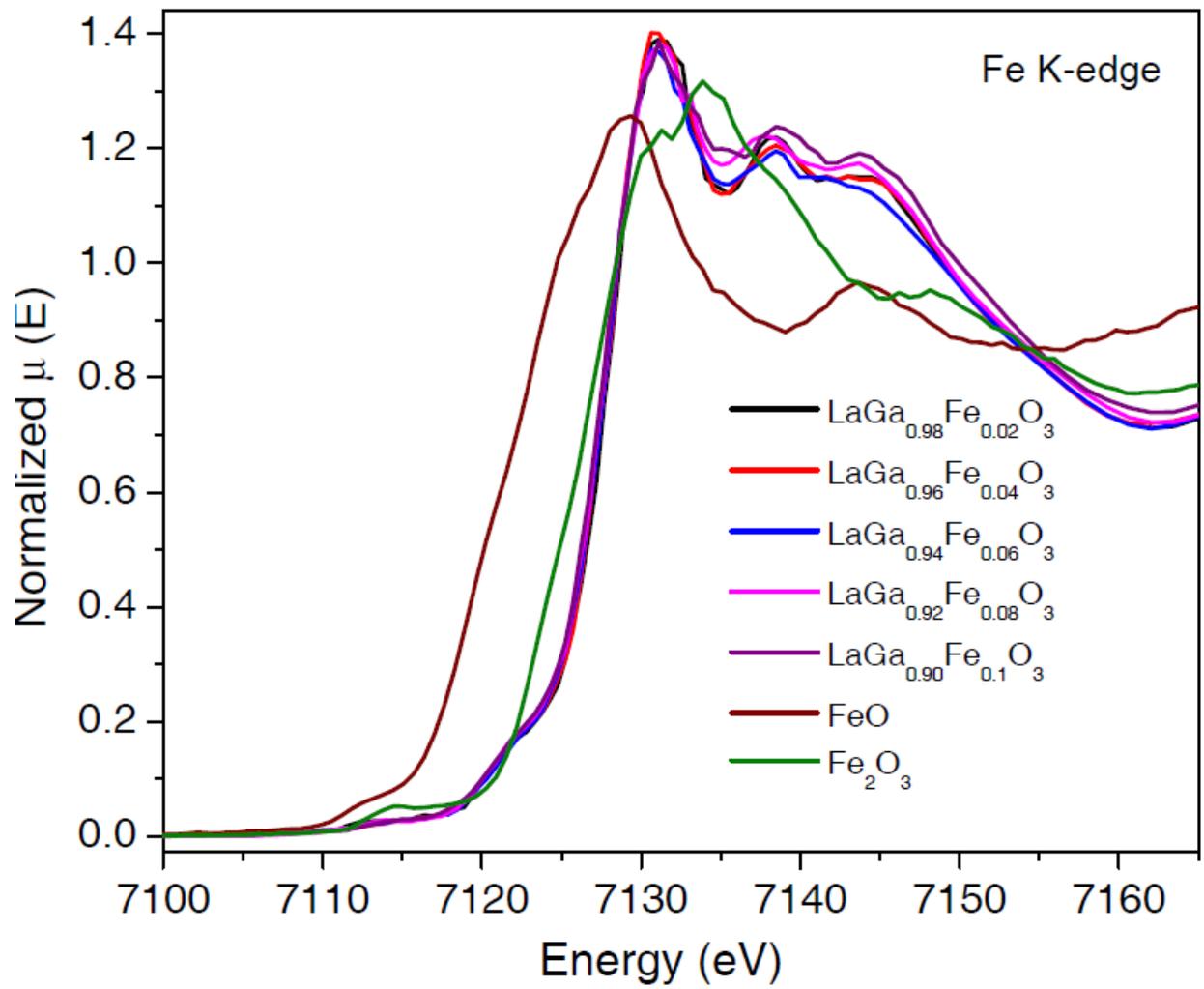

# Figure 6

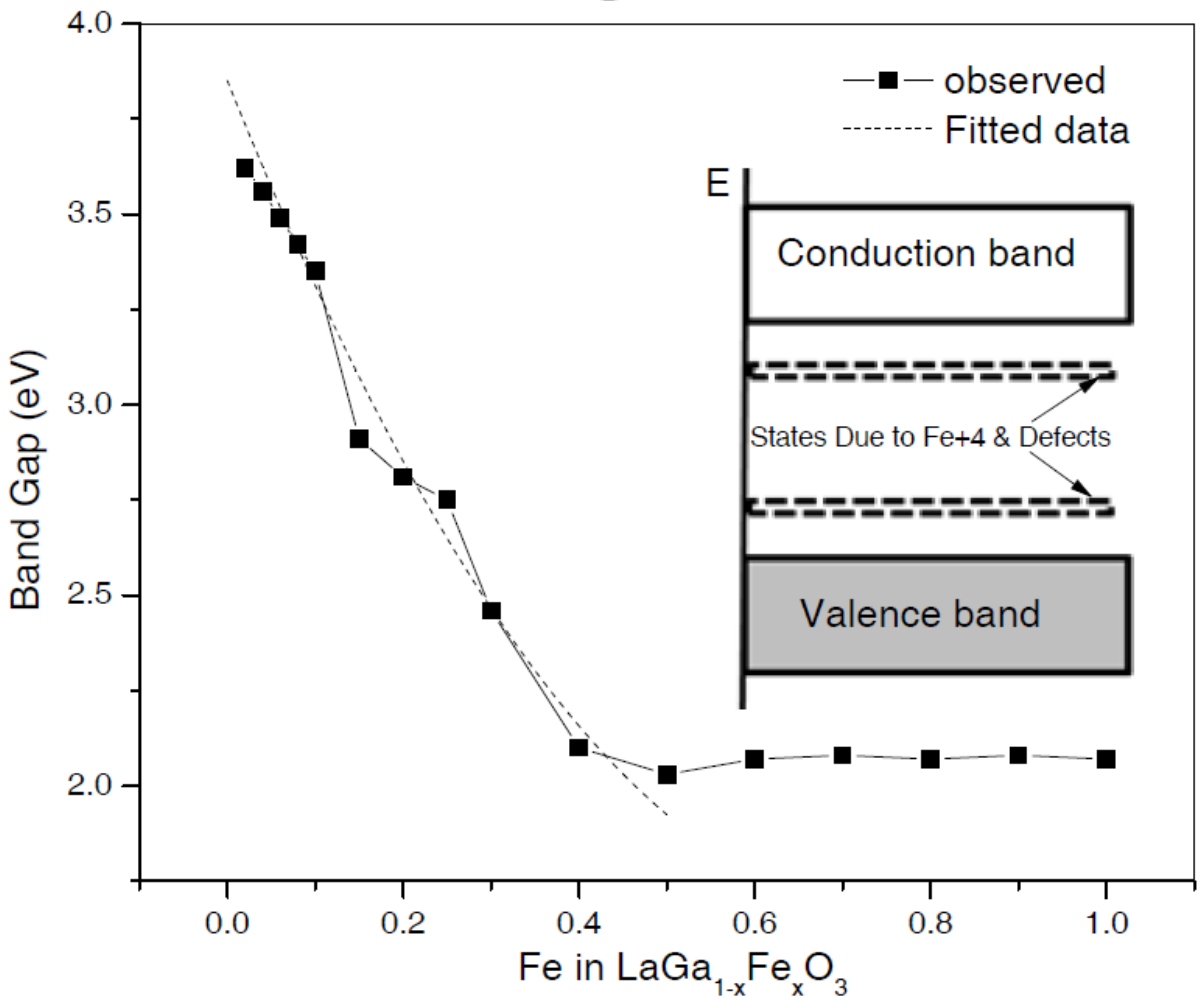